\documentclass{article}

\usepackage{cite}
\usepackage{amsmath,amssymb,amsfonts}
\usepackage{algorithmic}
\usepackage{graphicx}
\usepackage{textcomp}
\usepackage{xcolor}
\usepackage{verbatim} 
\usepackage{tikz}
\usepackage{amsmath}
\usepackage{url}
\usepackage{fullpage}
\usepackage{enumitem}
\usepackage{lipsum}
\usepackage{pgf-pie}
\usepackage{pgfplots}
\newcommand\df{Deepfake }
\usepackage{balance}

\usepackage{amsthm}

\newtheorem{objective}{Objective}[section]

\newcommand{\mycomment}[1]{}

\begin{document}

\title{Can deepfakes be created by novice users?}
\author{
  Pulak Mehta\\
  \texttt{pm3052@nyu.edu}
  \and
  Gauri Jagatap\\
  \texttt{gauri.jagatap@nyu.edu}
  \and
  Kevin Gallagher\\
  \texttt{k.gallagher@fct.unl.pt}
  \and
  Brian Timmerman\\
  \texttt{brian.timmerman@nyu.edu}
  \and
  Progga Deb\\
  \texttt{pd1353@nyu.edu}
  \and
  Siddharth Garg\\
  \texttt{sg175@nyu.edu}
  \and
  Rachel Greenstadt\\
  \texttt{greenstadt@nyu.edu}
  \and
   Brendan Dolan-Gavitt\\
  \texttt{brendandg@nyu.edu}
}
\date{}
\maketitle

\begin{abstract}
Recent advancements in machine learning and computer vision have led to the proliferation of Deepfakes. As technology democratizes over time, there is an increasing fear that novice users can create Deepfakes, to discredit others and undermine public discourse. In this paper, we conduct user studies to understand whether participants with advanced computer skills and varying level of computer science expertise can create Deepfakes of a person saying a target statement using limited media files.
We conduct two studies; in the first study ($n=39$) participants try creating a target Deepfake in a constrained time frame using \textit{any} tool they desire. In the second study ($n=29$) participants use \textit{pre-specified} deep learning based tools to create the same Deepfake. We find that for the first study, $23.1\%$ of the participants successfully created complete Deepfakes with audio and video, whereas for the second user study, $58.6\%$ of the participants were successful in stitching target speech to the target video. We further use Deepfake detection software tools as well as human examiner-based analysis, to classify the successfully generated Deepfake outputs as fake,  suspicious, or real. The software detector classified $80\%$ of the Deepfakes as fake, whereas the human examiners classified $100\%$ of the videos as fake. We conclude that creating Deepfakes is a simple enough task for a novice user given adequate tools and time; however, the resulting Deepfakes are not sufficiently real-looking and are unable to completely fool detection software as well as human examiners.
\end{abstract}

\section{Introduction} \label{sec:intro}

Deepfakes have gained attention both in technical literature as well as media. The concept of Deepfakes combines emerging techniques in \textit{deep} learning to create \textit{fake} multimedia content. It is a form of manipulated image, video, or audio data which can potentially be created with malicious intent.
Deepfake content is typically created with the following two objectives i) to superimpose a target person's video onto a video of a source person and mimic its actions and ii) to make the target person emulate audio from the source person. 

Pre-existing methods for generating artificial videos, use visual effects or computer graphics animation tools and require significant domain expertise to implement \cite{masson1999cg}. On the other hand, Deepfake creation tools have become easily accessible and require limited technical knowledge of the creation of fake multi-media content. The most popular platforms used are FaceSwap \cite{faceswap}, ZAO \cite{antoniou2019zao} and Reface \cite{reface} among several others. 
DeepFakesLab, another Deepfake open-source software, provides a comprehensive package \cite{perov2020deepfacelab}  for implementing several video manipulations such as face swaps, face aging, superimposing fake audio alongside lip syncing. FaceForensics \cite{rossler2018faceforensics, rossler2019faceforensics++} provides an open-source Deepfake image dataset.

Deepfakes are typically created using generative neural 
networks like Variational Autoencoders (VAE) \cite{kingma2013auto},
Generative Adversarial Networks (GAN) \cite{goodfellow2014generative} or flow-based generative models \cite{kingma2018glow}. 
The model is trained to explicitly learn the probability distribution of the image (or audio) training dataset. 
This allows the network to generate \textit{unseen} images (or audio) which have characteristics similar to the images (or audio) in the training set. Due to the abundance of image and audio data available online these days, particularly for images of popular actors, politicians and entrepreneurs, it has become increasingly easier to train  Deepfake generation models and produce realistic fake audio as well as videos. 

\subsection{Motivation} \label{subsec:motivate}

Deepfake's potential to deceive an average user gives it an edge over traditional image rendering. 
The Google search metrics \cite{gsm} for the term ``Deepfake" suggests a continuing upward trend in terms of interest among the general public. With the democratization of this technology, there is an imminent question: \textit{can ordinary citizens generate Deepfakes with malicious intent in a very short time frame?} 
We further motivate this question as follows:


\subsubsection{Accessibility} \label{subsec:access}

Platforms such as Faceswap \cite{faceswap} have given access to a common person with no domain skills to develop fake videos. Most applications \cite{thies2016face2face,ping2018deep} allow anyone with a webcam and microphone to produce a video with the target person emulating the expressions, lip movements and speech of the source speaker.  Along with the increasing processing power of GPUs and decreasing cost of cloud computing, these tools are becoming cost affordable \cite{collins2019forged}.

\subsubsection{Privacy and security} \label{subsubsec:privacy}
Over the past few years, social media and video-sharing platforms have been used with malicious intent to spread misinformation by impersonation of influential personalities. Deepfake videos targeting political figures include that of former United States (US) President Barack Obama \cite{forbes} using an expletive to describe the then US President Trump. 
Similar examples have since surfaced online, without being marked as fake videos. 


Several real videos have also been deemed as Deepfake, as a cover-up or deflection from political issues. When Gabon's president Ali Bongo \cite{deeptrace} released a video addressing the public after months of having not been seen in public, Bongo's political opponents deemed the video as a Deepfake, leading to increased distrust among common public. There is a significant social impact of people viewing videos without knowing if they are fake. 


Another area of risk is biometric authentication systems, like face authentication which has applications across Know-Your-Customer (KYC) compliance and fraud prevention. A recent study found that targetted deepfake attacks on face recognition tools achieved a success rate of $78\%$ \cite{10.1145/3485447.3512212} 

\subsubsection{Curating trustworthy datasets} \label{subsubsec:dataset}
Generative models, which are the backbone of Deepfakes, have also been used to augment datasets for training machine learning models wherever there is insufficient training data \cite{antoniou2018augmenting}. Deepfake detection has a direct implication in curating trustworthy datasets. 

\subsection{Social impact} \label{subsec:socialimpact}
The ease and quality of Deepfake video generation point to several social challenges.
Several papers have surveyed participants \cite{vaccari2020deepfakes,dobber2021microtargeted} and evaluated their reactions to Deepfake news articles. They conclude that Deepfake news articles, particularly when combined with political microtargetting techniques reduced participants' trust in social media news outlets and also changed their perception of the target person involved. This can have serious implications ranging from outcomes of elections to inducing distrust in governments and the economy. There has been a notable amount of media coverage \cite{guardian, forbes} on the threat to democracy that Deepfakes may pose.

Due to the threat that Deepfakes may pose, several government administrations  \cite{law} have come up with laws that enforce regulations on Deepfakes, including the Deepfake Report Act under the 2021 National Defense Authorization Act (NDAA) of the US government. US state governments of Texas and California have also passed laws banning use of Deepfakes close to elections \cite{law}. 

\subsection{Limitations of current Deepfake platforms} \label{subsec:limit}

It is important to note that the most easily available mobile applications such as FaceSwap \cite{faceswap}, ZAO \cite{antoniou2019zao}, and Reface, have limited use cases.
They are targeted towards generating video fakes with fixed audio files added onto the target video. These audio files may be pre-recorded clips from songs or movie dialogues and are only created for the purpose of entertainment. 
These videos generally conform to the policies of major social platforms and are not perceived as a risk to the society. 


Through our discussion above, we note that most Deepfake content has high profile personalities. In this paper, we study a more challenging problem -- that of creating a Deepfake with an ordinary person as the target. We then assess the risk it may pose, by using human and software detection tools.

In the next section,  we outline the main questions addressed in this paper, as well as summarize key insights.

\subsection{Our contributions} \label{subsec:contri}

The existence of Deepfakes lead to several
pertinent questions that we address in this paper. In particular, we probe the following
\begin{enumerate}
    \item Are Deepfakes easy to generate for a {novice user}\footnote{By ``novice" we refer to users who do not have direct expertise in creating Deepfakes.} with varying skill levels?
    
    \item For the Deepfakes generated by our sample participants, are these Deepfakes generated realistic enough to fool
    \begin{enumerate}
          \item a set of human examiners
        \item a Deepfake detection software tool?
    \end{enumerate}
\end{enumerate}

We aim to answer these questions via our structured methodology that we  highlight in detail in Sec. \ref{sec:methodology}. Particularly, we design an objective where we pick both the source and target person. We assign two objectives to all categories of participants selected: 

\begin{objective} \label{obj:1}
(video) to transfer the expressions and lip movements of source speaker to that of target speaker
\end{objective}

\begin{objective} \label{obj:2}
(audio) to transfer the speech of source speaker to that of target speaker.
\end{objective}
We conduct a two-fold study for the same; in the first part, we assign an open-ended objective and ask participants to use \textit{any} Deepfake generating tool of their choice. In the second part, we ask the participants to use a \textit{pre-defined} tools to create Deepfakes. We deem the task of creating a Deepfake as successful if the participants are able to synthesize a video with the target speaker uttering the target speech. Our study yields the following insights summarized below:

\begin{enumerate}
\item Deepfake generation:

\begin{enumerate}
    \item $ 23.1\%$ of participants are able to complete both the Deepfakes Objectives \ref{obj:1} and \ref{obj:2} in the open-ended study.
    \item $58.6\%$ of participants are able to complete both the Deepfakes Objectives \ref{obj:1} and \ref{obj:2} successfully in the pre-defined tool study
\end{enumerate}
    
    \item Deepfake detection: Deepfakes generated by participants are flagged as fake or suspicious 
    \begin{enumerate}
        \item for $100\%$ of Deepfake samples, when assessed by a set of human examiners.
        \item  for $ 80\%$ of Deepfake samples, when analyzed by a software detection tool. 
    \end{enumerate}
\end{enumerate}

In the next section, we provide a comprehensive literature survey of topics relevant to our study.

\section{Paper organization} \label{sec:paperorganize}
We have organized the paper as follows. In Sec. \ref{sec:related} (Related work) we discuss survey literature in related areas of Deepfake generation techniques (Sec. \ref{subsec:dfgen}), Deepfake detection techniques (Sec. \ref{subsec:dfdet}), User studies on Deepfake detection (Sec. \ref{subsec:userdet}) and User studies on Deepfake generation (Sec. \ref{subsec:usergen}). In Sec. \ref{sec:methodology} (Methodology) we outline the main structure of our two user studies; in particular we discuss in Sec. \ref{subsec:userstudy}, details on participant background in Sec. \ref{subsec:partback} and participant recruitment in Sec. \ref{subsec:recruit}. We finally discuss the process of conducting the user study in Sec. \ref{subsec:studyproc}.

In Sec. \ref{sec:find} we focus on the main findings of our user studies. We comment on the participant awareness about Deepfakes in Sec. \ref{subsec:awaredf}. This is followed by main findings from our Open-ended Deepfake generation Sec. \ref{subsec:opedfstudy}, Pre-defined Deepfake generation study Sec. \ref{subsec:predfstudy}. We finally present our Discussions and Conclusions in Sec. \ref{sec:disussions}.

In Appendix \ref{sec:appendix}, we elaborate deep generative literature, details on participant compensation and demographics, ethical considerations, limitations of our approach and future direction. We also include some snapshot examples from deepfakes created in Appendix \ref{sec:supplementary}.

\section{Related work} \label{sec:related}

We push the review on generative networks to Appendix \ref{sec:appendix}. 

\subsection{Targeted Deepfake generation} \label{subsec:dfgen}

One of the first attempts at targeted Deepfake generation is \cite{suwajanakorn2017synthesizing}, where the authors synthesize a high-quality lip-synced video of President Barack Obama given an input audio clip. The model was trained on nearly two million frames of Obama's weekly address footage using a recurrent neural network (RNN) that maps raw audio features to mouth shapes. 

Face2Face \cite{thies2016face2face} is another important benchmark and one of the first methods to successfully render fake facial expressions and audio, using a dense photometric consistency measure. 

FaceSwap\cite{faceswap} and DeepFaceLab\cite{perov2020deepfacelab} are both open source software for Deepfake generation. Wav2lip \cite{prajwal2020lip} provides automatic lip-syncing given a target video and corresponding audio to be synthesized. It uses a pre-trained Discriminator, referred to as a ``lip-sync expert" which allows their network to produce \df video with high fidelity between audio and lip movements. 

Another approach is first-order motion \cite{siarohin2019first}, where the Deepfake generation process is split into two parts; motion extraction via unsupervised key point detector, and generation which in-paints the target video along the key points extracted.

Some other recent papers include FSGAN, which does not require any training on the source or target subjects \cite{nirkin2019fsgan}, RSGAN \cite{natsume2018rsgan} which allows additional flexibility in explicit feature editing (like controlling the appearance of hair or eyes separately) and FaceShifter \cite{li2019faceshifter} which improves the fidelity of generated \df video under facial occlusions. 

It is worth noting that the code for implementing most of these Deepfakes are \cite{prajwal2020lip}, \cite{nirkin2019fsgan}, \cite{natsume2018rsgan} open-source and requires the user to just have access to a computer with basic hardware requirements. Platforms like Google Colaboratory also provide free access to limited GPU computing.

\subsection{Deepfake detection techniques} \label{subsec:dfdet}

The Deep Fake Detection Challenge (DFDC) \cite{dolhansky2020deepfake} 
consists of $\sim124,000$ videos and featured eight facial modification algorithms. The winning detection technique from the challenge by Seferbekov \cite{seferbekov} consists of breaking the video down frame by frame and using a multi-task cascaded convectional network (MTCNN) for face detection \cite{zhang2016joint} and extracting a crop that excludes the background. This is followed by running an EfficientNet \cite{tan2019efficientnet} classifier to detect the face crop as real or fake.

Deepware \cite{deepware} is an API that directly takes a video (fake or real) from the user and tests against four different machine detection algorithms. The four detectors considered are: i) avatarify \cite{avatarify} which is a face animation app, ii) deepware \cite{deepware}, iii) the MTCNN and EfficientNet methodology by Seferbekov \cite{seferbekov}, and iv) an ensemble method that combines deepware detector and Seferbekov's detector. Other techniques for detection that are not covered by Deepware API include but are not limited to: Mesonet, which utilizes mesoscopic image information to build a neural classifier \cite{afchar2018mesonet} and Face-Cutout \cite{das2021towards} which is a data-augmentation procedure which improves upon \df detection accuracy of prior art.

We refer the reader to \cite{nguyen2019deep,mirsky2021creation,tolosana2020deepfakes} for a more comprehensive overview of both Deepfake generation and detection techniques.

\subsection{User studies on Deepfake detection} \label{subsec:userdet}

In academia, there is no clear consensus on whether human participants can successfully distinguish between real and fake videos. 

\cite{tahir2021seeing} provides a comprehensive study on human participants equipped with the task to assess if provided Deepfakes were real or fake. Their survey suggests that majority of the people are unable to classify a \df image to be fake. 

The study in \cite{kobis2021fooled} suggests that participants had a systematic bias toward guessing that videos are authentic, fake or real. They conjecture that people apply an overly optimistic seeing-is-believing heuristic, which points to a larger risk of the common public being influenced by Deepfakes. 

The study in \cite{groh2022deepfake} suggests that \df detection software and human examiners have similar classification accuracies on the same dataset, however, make different mistakes when misclassifying a media. They propose that a combined classification procedure that uses both machine detection and human visual examination is the best strategy when it comes to \df detection. 

We refer the reader to \cite{mangaokar2021dispelling} which characterizes the shortcomings of the current state of \df detection techniques. 

\subsection{User studies on Deepfake generation} \label{subsec:usergen}

To our knowledge, there is no other paper that discusses the capability of novice users to make Deepfakes. We now present the main setup of our user study in the next section. 


\section{Methodology} \label{sec:methodology}

\begin{figure*}
\begin{center}
\includegraphics[width=0.7\textwidth]{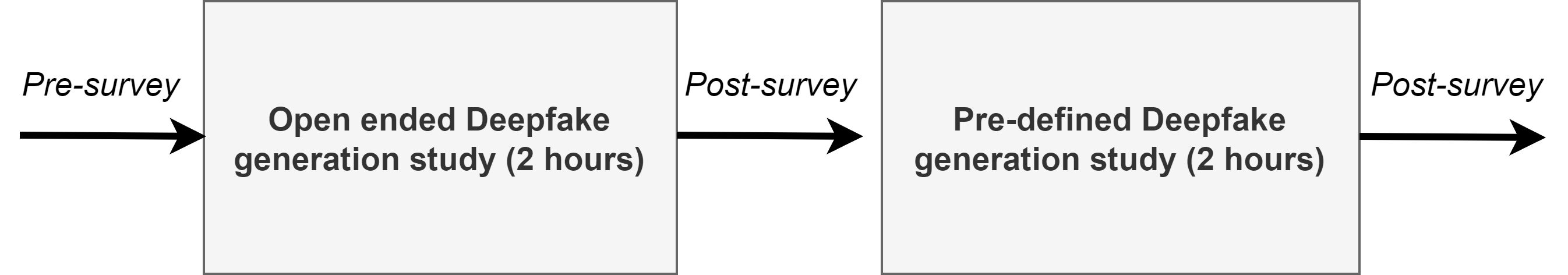}
\caption{Overview of the two-part user study. The pre-survey captured participants' awareness of Deepfakes.}\label{fig:fig1}
\vspace{5pt}
\end{center} 
\end{figure*}

We start with outlining the various considerations that went into designing and conducting the two-part user study on Deepfake generation. See Figure \ref{fig:fig1} for a visual representation of the study. 

\vspace{5pt}
\subsection{User study} \label{subsec:userstudy}
 We devise a user study to understand Deepfake generation by requiring a set of pre-selected participants to create an audio-visual Deepfake of a target person speaking a target statement. This is the most general case of Deepfake generation and can have major social-economical ramifications. In our study, the target person was Kevin who is a project member and the target statement was:
 
 \emph{``Hello everyone, my name is {Kevin} and this is a Deepfake video".}

Standards techniques for Deepfake generation techniques rely on transferring the lip and facial movements from one video clip onto another video. We provide the following media files to the {novice user} to create a Deepfake: (i) video of the target person and (ii) a video clip of source speaker speaking the statement that the target needs to emulate.

The task is assessed as successful if the output of the fake audio-visual used a Deepfake generation tool to create the audio and video components along with stitching them together. 

The participants were provided with three videos, each being $\sim1$ hour long, of the target person (target video style) speaking in their voice (target audio style). Additionally, they were provided with a video clip of another research participant (source video) speaking a $\sim$6 second clip of the target statement in their voice (source audio). 

We conducted four pilot studies to iterate and determine the best structure for the user study. These pilots consisted of two participants from the research team who have experience with deep machine learning techniques and two participants with limited machine learning knowledge but from a computer science background. The pilot participants were asked to create Deepfakes on a whim with the same media files. Three of the participants were able to complete the task within 1.5 hours with no guidance, while the last participant finished within 2 hours with limited guidance on debugging the Deepfake creation tool. The biggest challenge faced by the pilot participants was identifying the right tool, followed by debugging while using these tools. 

Based on this, considerations for designing the Deepfake generation user study are as follows:
\subsubsection{Time frame} \label{subsec:timeframe}
In the pilot study three out of four participants needed less than 1.5 hours implying that generating Deepfakes in a time-bound manner is possible. Moreover, we also want to assess how practical it is to make Deepfakes \textit{on a whim}. Studies \cite{satputedeepfake} suggest that certain Deepfakes can be created in less than 6 minutes. Based on these considerations, we assign a time limit of 2 hours for both parts of the study.

\subsubsection{Ease of accessibility of Deepfake generation tools} \label{subsec:dftoolacc}
We investigate if the identification and accessibility of the Deepfake generation tool is a major challenge in the creation of Deepfake content. We design a two part study, one where tools for creating Deepfakes are not provided, and one where they are. 

\subsubsection{Ease of implementation of \df tools}  \label{subsec:dftoolimpl}
We construct a survey to assess the difficulties experienced while implementing \df algorithms. 

These factors lead us to create the two parts of the study. 
 \color{black}

The first part of the study required the participants to create Deepfake using \textit{any tool}. We refer to this part of the study as \emph{Open-ended Deepfake generation study} for the remainder of this paper. 

For this part of the study, participants are free to use search engines, mobile applications, online software tool, or reuse any code base. They are allowed to use their personal computers or available cloud servers to generate the desired fake video. The study is designed to be representative of Deepfake creation in the real world. To set a control parameter for the study, we allot all participants 2 hours for the first part of the study. 
 
In the second study, we re-invite the same participants for creating the same target Deepfake using \textit{pre-specified tools} for fake audio and video generation in a time frame of 2 hours. We subsequently refer to this part of the study as \emph{Pre-defined Deepfake Generation study}. The participants are also provided with publicly available tutorials for using these tools. More details about the tools are provided Sec. \ref{sec:pre-define}. 
 
Using this two-fold approach, we investigate whether Deepfake generation tool identification is a major challenge in the generation of Deepfakes. The second part of our study enables us to examine if given the correct set of tools, can {novice users} generate Deepfakes. 

Lastly, we run both software-based and human-examination-based detection on the fakes generated to investigate how many of the videos generated are actually \textit{realistic}. 

\subsection{Participant background} \label{subsec:partback}
Generation of audio or video Deepfakes from scratch requires skills such as software, machine learning and basic media editing like cropping, merging media etc. Additionally, video Deepfakes can be enhanced using Visual Effects (VFX) techniques \cite{9576757}. Hence, for our study, we select four categories of participants that represent the broad skill sets required to create fakes. These categories are- 
\begin{enumerate}
  \item Basic computer skills - The only coding coursework  completed by the participant is introduction level courses on computer science, during their academic coursework.
  \item Intermediate computer skills - Participants have studied multiple computer science or coding courses in their coursework, but not machine learning.
  
  \item Advance computer skills - Participants have taken at least one machine course during their coursework.
  \item Digital Media skills - They have completed at least one coursework related to Visual Effects.
\end{enumerate}

All participants surveyed have a basic understanding of a computer either through an introduction to computer science course or a Visual Effects class and possess the ability to perform basic media file edits, such as cropping, merging etc. or can quickly grasp the concepts for the same during the experiment session.
 We received a few participants with overlapping categories of basic computer skills and digital media skills. They were placed under digital media skills for the study.

\subsection{Recruitment} \label{subsec:recruit}

Recruitment emails were sent to the first author's university mailing list stating that this study is regarding Deepfakes. The email did not contain any information about the actual tasks that the participants would be required to do for the study. To sign-up for the study, participants provided their details such age, degree program enrolled, relevant coursework, gender, and ethnicity  along with their consent for the voluntary study. We attempted to select a diverse set of participants across each of the participant categories. We expected dropouts as the study had multiple parts.
 Hence we invited 52 participants ($13 \times 4$ categories) for the study. Out of the 52 participants, only 39 turned out for the open-ended tool study, and 29 for the pre-defined tool study. The breakdown of participants that were finalized is described in Appendix \ref{subsec:partdemo}.

\subsection{Study procedure} \label{subsec:studyproc}

The sessions were conducted virtually using Zoom in compliance with the health safety advisor by the local authorities. At the start of a session, participants switched on their webcams for the duration of the sessions and were allotted a unique identification code which remained the same for both of the studies. They were required to change their display name to the identification code.

For the first session, a pre-survey was conducted to understand their knowledge of Deepfakes. After the pre-survey, participants were provided with the task for that session. All participants attempted the Deepfake creation task in the allotted time of the corresponding study. During each study, we provided assistance through a stop point survey (see Figure \ref{fig:fig1}). If participants faced any challenge for more than 10 min and required guidance for the next step, they were instructed to fill out a brief stop-point survey highlighting their problem. A member of the research team was then assigned to guide them in solving the problem in a breakout room. The research team only provided direction and did not debug the code or assist in solving the problem completely. After the allocated time, the participants filled out a brief post-survey alongside submitting their outputs. 

In Table \ref{tab:survey1} and Table \ref{tab:survey2} we break down the survey rubric for the pre-study survey and post-study survey respectively.

\begin{table}[!t]
    \centering
        \caption{Breakdown of survey rubric prior to user study.}
    \begin{tabular}{|l|l|}
    \hline 
    Rubric & Questions \\
    \hline 
        Awareness &  Heard about Deepfakes prior? \\ & Sources of information?\\
        \hline 
        Knowledge & How Knowledgeable about Deepfakes? \\
        \hline 
        Expertise & Have they created Deepfakes before?\\
        & How optimistic about finishing the objective? \\
        \hline 
    \end{tabular}

    \label{tab:survey1}
\end{table}

\begin{table}[!t]
    \centering
     \caption{Breakdown of survey rubric post user study.}
    \begin{tabular}{|l|l|}
    \hline 
    Rubric & Questions \\
    \hline 
        Techniques & Broad strategy to solve objective?
        \\
        & Most challenging step?\\
        & Time spent on each step?\\
        & Software tools used? \\
        \hline 
        Confidence &  How optimistic about creating Deepfakes?\\
        & Did they lose motivation during study?\\
        \hline 
        Task  & Fake video created?\\
        completion & Fake audio created?\\
        & Fake audio and video combined?\\
        & Deepfake  video realistic? \\
        \hline 
    \end{tabular}
   
    \label{tab:survey2}
\end{table}

We describe details about participant compensation and ethical considerations in Appendix \ref{sec:appendix}.

\section{Findings} \label{sec:find}
\subsection{Awareness around Deepfakes} \label{subsec:awaredf}
\begin{figure}[!h]
\begin{center}
\begin{tikzpicture}[scale=0.8]
\pie[rotate=15,radius=1.7, 
color={blue!30,green!30,red!30}
]{84.6/Known previously,
    10.3/Recruitment email,
    5.1/Never heard}
\end{tikzpicture}
\caption{Deepfake awareness among participants.} \label{fig:fig2}
\end{center}
\vspace{-0.3cm}
\end{figure}
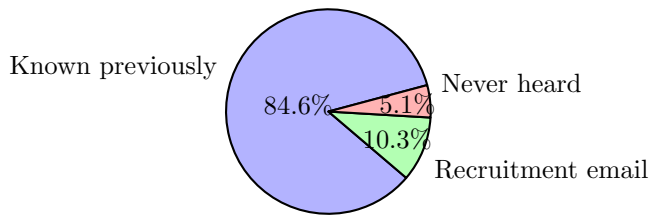

We collected insights about Deepfakes awareness from the pre-survey (Table \ref{tab:survey1}) at the start of the open-ended Deepfake generation study. Majority of the participants were aware of Deepfakes ($\sim$84.6\% ) before the recruitment email, see Figure \ref{fig:fig2}. These participants reported having heard about Deepfakes across various platforms. The most popular platforms were social media sites (n=23), closely followed by online video sites (n=22), news outlets (n=19), friends and family (n=17), academic work (n=14), and messaging apps (n=4) in decreasing order. The most common social media website where participants encountered Deepfakes was Instagram and TikTok, while the most common online video site was Youtube.

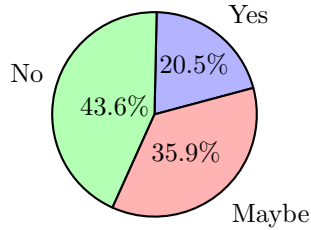
\begin{figure}[!h]
\begin{center}
\begin{tikzpicture}[scale=0.8]
\pie[rotate=15,radius=1.7, 
color={blue!30,green!30,red!30}
]{20.5/Yes,
    43.6/No,
    35.9/Maybe}

\end{tikzpicture}
\caption{Did the participants see Deepfakes \textit{without} any disclaimer?}\label{fig:fig3}
\end{center} 
\end{figure}

Additionally, a small number of participants ($\sim$20.5\% ) reported having seen Deepfakes on the internet without a Disclaimer stating that they were Deepfakes (see {Figure \ref{fig:fig3}}). This can be a cause of concern when Deepfakes, when misinformation targeted fake, is available online without disclaimers.

At the start of the study, only 3 out of the 39 participants ($\sim$7.7\%) had previously attempted to create Deepfakes using free mobile apps with the objective of creating memes/ humorous content.

\subsection{Open-ended Deepfake generation study } \label{subsec:opedfstudy}
\subsubsection{Overview} \label{subsubsec:overview}
In this session, participants were required to generate the target Deepfake in 2 hours without being recommended any specific tool or tutorial. Participants were provided all the media files (three of the target person and one of the source persons speaking the target statement). Out of the 41 participants who started the study, two of them dropped mid-way. One of them was worried that the study would require them to download software and they may accidentally download malicious software. A second participant dropped out due to unknown reasons. Overall 39 participants completed the session (a breakdown of the technical expertise of these participants is presented in Table \ref{tab:table1}). 

\begin{table}
\begin{center}
\caption{\label{tab:table1}Breakdown of technical expertise for participants in Open-ended Deepfake generation study. We ensure that number of participants in each category is roughly equal.}
\begin{tabular}{||c c ||} 
 \hline
 Category & Participants \\ [0.5ex] 
 \hline\hline
 Basic computer skills & 10 \\ 
 \hline
 Intermediate computer skills & 10 \\
 \hline
 Advance computer skills & 11\\
 \hline
 Visual Media Skills & 8 \\
 \hline
 \hline 
 Overall & 39 \\ [1ex] 
 \hline
\end{tabular}
\end{center}

\end{table}

\begin{table}
\begin{center}
\caption{\label{tab:table2}(Stop point survey from open ended study) Breakdown of the key challenges encountered to generate a \df. Most of the participants encountered challenge in debugging the code on Google Colab.}
\begin{tabular}{||l c ||} 
 \hline
 Challenge type & Percentage \\ [0.5ex] 
 \hline\hline
 Debug - Google Colab & \textbf{35.8} \\ 
 \hline
 General query & 17.9 \\
 \hline
 Debug - Local computer & 14.3\\
 \hline
 Tool identification & 14.3 \\
 \hline
 Tool taking time & 10.7 \\
 \hline
 Media editing & 7.1 \\ [1ex] 
 \hline
\end{tabular}
\end{center}

\end{table}

\subsubsection{Stop point survey} \label{subsubsec:stoppoint}
We received 28 requests for support with Deepfake generation using the stop-point survey. 50\% of the issues were related to generating video fakes, 35.7\% were related to generating audio fakes and the remaining were general issues like clarification of the study's objective.  Hence, even with limited or no knowledge of making fakes, most people were able to make headway in creating them. However, not all participants identified the correct set of tools to implement the task. Most of the participants faced challenges in debugging the identified tool. These predominately arose in debugging tools running on Google Colaboratory, a platform that allows users to run Python code. We highlight these findings in Table \ref{tab:table2}.

The motivation of participants also varied during the session. 60\% of the participants reported that they had sustained motivation throughout the study. The loss in the motivation of the other 40\% was attributed to factors such as long debug time, lack of coding or machine learning knowledge etc. Interestingly, some participants reported that they became more motivated as time progressed. In terms of time split while creating Deepfakes, $\sim$56.3\% of the time by participants was spend on video fakes, $\sim$26.4\% on audio fakes and $\sim$20.3\% on other activities like tool identification, media editing etc. 

\subsubsection{Deepfakes generated} \label{subsubsec:dfgenerated}

Within a two-hour period, a sizeable number of the participants were able to generate Deepfakes. The outcomes are highlighted in Table \ref{tab:table3}.

\begin{table}
\begin{center}
\caption{\label{tab:table3} (Output from open-ended study) Based on human review, we find that $23.1\%$ of participants are able to create a Deepfake. These numbers do not represent if the \df created can be detected as a fake by the human examiners or software detector.}
\begin{tabular}{||c c ||} 
 \hline
 Deepfake component created & Percentage \\ [0.5ex] 
 \hline\hline
 Deepfake Video without audio & 15.4\% \\ 
 \hline
 Deepfake Audio without video & 10.2\% \\
 \hline
 Deepfake audio and video & 23.1\% \\ 
 \hline 
 Neither audio nor video & 51.3\% \\
 \hline
\end{tabular}
\end{center}
\end{table}

Our criteria of whether a given participant created a Deepfake successfully, is whether they used an existing Deepfake tool to create the audio and video. This meaningful completion rate of 23.1\% can be attributed to Deepfake tools being made publicly available on Google Colab with detailed instructions by the authors. We add details about techniques used in Deepfake generation in the open-ended study as follows in Tables \ref{tab:method1} and \ref{tab:method2}. 
\begin{table}[!h]
    \centering   \caption{(Open-ended study). Breakdown of techniques used by 15 people who created fake video successfully. These numbers do not represent if the \df created can be detected as a fake by the human examiners or software detector.}
    \vspace{+0.3cm}
    \label{tab:method1}
    \resizebox{\columnwidth}{!}{
    \begin{tabular}{|c|c|c|}
    \hline 
    Method & Details & Num.\\ & & participants\\
    \hline

        First Order Motion\cite{siarohin2019first} & Academic open source & 5\\
        Speakr & Mobile app & 2\\
        Reface & Mobile app & 2\\
       DeepFakeLab \cite{perov2020deepfacelab} & Open source &  1 \\
       Wav2lip \cite{prajwal2020lip} & Academic open source & 1\\
        Adobe After effects & Image enhancement software & 1\\
        DaVinci & Image enhancement software & 1\\
        Snapchat  & Mobile app & 1\\
        Avatarify \cite{avatarify} & Web tool & 1\\
        \hline 
        Total& & 15\\
        \hline 
    \end{tabular}}

\end{table}

\begin{table}[]
    \centering
     \caption{(Open-ended study). Breakdown of techniques used by 13 people who created fake audio successfully. These numbers do not represent if the \df created can be detected as a fake by the human examiners or software detector.}
     \vspace{+0.3cm}
    \label{tab:method2}
    \begin{tabular}{|c|c|c|}
    \hline
        Method & Details & Num.\\
        & & participants \\
        \hline 
        
       SV2TTS \cite{jemine2019master} & Text-to speech open source & 9 \\ 
       DeScript & Web tool & 1\\
       Tacotron 2 \cite{shen2018natural} & Text-to speech open source & 1\\
        Mozilla TTS & Text-to speech open source & 1\\
        \hline 
        Total & & 13 \\
        \hline 
    \end{tabular}

\end{table}

After generation, the next step is to analyze if the Deepfake quality of the video is good enough to mislead humans or computers.

\subsubsection{Deepfakes detection} \label{subsubsec:dfdetection}
Most of the work around Deepfake detection revolves around detecting video, and not the audio component. We process all the videos through a Deepfake detector to identify whether it can bypass detectors. We leverage the  Deepfake detection API provided by Deepware\cite{deepware} to quantify the quality of the fake. Their tool uses four major algorithms to determine if video is fake or not - Seferbekov\cite{seferbekov}, Avatarify, Deepware and an ensemble method that leverages the power of the previous three methods. Seferbekov's method \cite{seferbekov} is the winner of the popular Deepfake Detection Challenge (DFDC)\\ \cite{dolhansky2020deepfake}, whereas Avatarify which is aimed at detecting fakes generated using the Avatarify tool. The output of the analysis for the Deepfakes created in the open-ended study is provided in Table \ref{tab:table4}.

\begin{table}
\begin{center}
\caption{\label{tab:table4} Of the 48.7\% videos that were partially or completely generated as Deepfakes, we use the Deepware \cite{deepware} software tool to flag the videos as real or fake. We find that most of the \df videos created by participants are flagged as fake using Deepware.}
\vspace{+0.3cm}
\begin{tabular}{||c c ||} 
 \hline
 Video type & Percentage \\ [0.5ex] 
 \hline\hline
 Detected fake & \textbf{53.3\%} \\ 
 \hline
 Detected suspicious & 26.7\% \\
 \hline
 Not detected & 20\% \\ [1ex] 
 \hline
\end{tabular}
\end{center}

\end{table}

The Deepware API tool processes video and each of the four algorithms reports a score from 0-100, where 100 is fake. The suspicious videos had an average score algorithmic score between 52-72 by each of the four algorithms. The videos that were flagged as ``not fake" video had the best algorithm report a score of 48 and 49. There are many factors that we observed that determined the fakeness by this detector. This included the number of faces in a video, face movement, face direction etc. An interesting observation was that one student from the digital media background used DaVinci resolve, which is a video editing application, to manipulate the lip movements. Such a process is extremely time-consuming to create a realistic fake, however, the best out of the four algorithms provided it with a score of 7. Hence, future detectors should be trained to identify fakes that can be generated using video editing software like DaVinci Resolve.

We further conducted a human-based analysis of the videos generated. It consisted of five examiners of the following background - \\
1) 20 year old, male, social science background \\
2) 23 year old, female, physical science background \\
3) 24 year old, male, physical science background \\
4) 27 year old, female, journalism background \\
5) 29 year old, female, engineering background \\
The human analyzers gave us their consent; however were not financially compensated. They reviewed the outputs from the open-ended study on a laptop and all values reported by this group hereafter have been averaged out. Note that the users were informed before-hand that the videos they view may be real or fake.

We instruct the examiners as follows:
\begin{enumerate}
    \item detect if the video with muted audio is fake or real
    \begin{enumerate}
        \item if detected as fake, comment on which parts of the video (including facial features) has distortions.
    \end{enumerate}
    \item detect if the audio without the video sounds 
    \begin{enumerate}
        \item like human
        \item vaguely like target
        \item similar to target
    \end{enumerate}
    \item detect if audio and video played together looked
    \begin{enumerate}
        \item lip-synced
        \item realistic; i.e. no video distortions; sounds similar to target.
    \end{enumerate}
\end{enumerate}
The reviewers reported only $\sim~$ 25.6\% of the videos had the face looking like target person. The other videos included superimposing select features of source like eyes, mouth onto Kevin's hair and beard, massive distortion etc. Such details may bypass a programmatic detector but can be easily identified by the human eye.

Analysis of human examination of the Deepfakes generated in the open-ended study is in Table \ref{tab:table5}. Similar to \cite{tahir2021seeing}, we dissect the distortions in different zones of the video, such as eyes, lips, etc. Only found two videos were flagged as real with muted audio.

\begin{table}[!h]
\begin{center}
\vspace{-0.1cm}
\caption{\label{tab:table5}Human expert based visual inspection of properties of Deepfake video that appear unreal and may cause the \df video to be flagged as fake. In most \df videos we find that the lips/mouth movement to be the most distorted (highlighted).}
\vspace{+1cm}
\begin{tabular}{||c c ||} 
 \hline
 Video Area & Videos not Distorted \\ [0.5ex] 
 \hline\hline
 \hline
 Eyes & 82.05 \% \\
 \hline
 Facial hair & 76.92 \% \\
 \hline
 Lips/Mouth & \textbf{69.23} \% \\
 \hline
 Nose & 74.36 \% \\
 \hline
 Cheeks & 71.79 \% \\
 \hline
 Forehead & 74.36 \% \\
 \hline
 Hair & 71.79 \% \\
 \hline
 Background/other & 82.05 \% \\
\hline
\end{tabular}
\end{center}
\end{table}

We also conducted an audio analysis of these fakes by human reviewers. The output of most fakes sounded like human. Our analysis of the audio clips with no video is highlighted in Table \ref{tab:table6}.

\begin{table}[!h]
\begin{center}
\caption{\label{tab:table6} Human based auditory inspection of Deepfake audio quality. Most \df videos have audio that sounds like a human, but not necessarily like target person.}
\vspace{+1cm}
\begin{tabular}{||c c ||} 
 \hline
 Audio Details & Percentage \\ [0.5ex] 
 \hline\hline
 \hline
 Sounds like human & 65.24\% \\
 \hline
 Sounds vaguely like target  & 34.76 \% \\
 \hline
 Sounds similar to target & 0.0 \% \\
\hline
\end{tabular}
\end{center}
\vspace{-0.5cm}
\end{table}

Upon watching the audio and video together we found only 15\% of the videos to be vaguely lip-synced. However, none of the fakes were realistic enough to demonstrate that the target person was speaking the target statement.

\subsection{Pre-defined Deepfake generation study } \label{subsec:predfstudy}

\subsubsection{Overview} \label{sec:pre-define}
In this session participants were asked to generate the target Deepfake in 2 hours using a specific tool. The tools that we provided for creating fakes were the Google Colaboratory version for the Real-Time Voice Cloning \cite{jemine2019master} and Wav2lip \cite{prajwal2020lip} for audio and video respectively, which are state-of-art open source repositories. 
We provided access to condensed tutorials for the audio \cite{audiotutorial} and video \cite{videotutorial} based on reference papers \cite{prajwal2020lip,jemine2019master}. Additionally, the Google Colaboratory version of these tools ensured that all participants had access to similar hardware.

All participants were given the same media files as the prior session. Due to dropouts, only 29 completed this study (breakdown of their technical expertise is presented in the Table \ref{tab:table7}). 

\begin{table}
\vspace{0.4cm}
\begin{center}
\caption{\label{tab:table7}Breakdown of technical expertise for participants in the pre-defined Deepfake generation study. }
\vspace{+1cm}
\begin{tabular}{||c c ||} 
 \hline
 Category & Participants \\ [0.5ex] 
 \hline\hline
 Basic computer skills & 7 \\ 
 \hline
 Intermediate computer skills & 9 \\
 \hline
 Advance computer skills & 7\\
 \hline
 Visual Media Skills & 6 \\
 \hline
 \textbf{Overall} & \textbf{29} \\ [1ex] 
 \hline
\end{tabular}
\end{center}

\end{table}

\subsubsection{Stop point survey} \label{subsubsec:stoppointstudy2}
We received 15 requests for support during the study which we captured through the stop-point survey. Since the tools and tutorials were provided most issues were either related to debugging Google Colaboratory (10), general queries (3), or editing media (2). We observed that Safari browsers, even with the latest version, were facing problems but as soon as they switched to Google chrome their problem got resolved.  

\subsubsection{Deepfake generated} \label{subsubsec:dfgenerated2}
To generate the Deepfake, participants first created the audio fake using the Real-Time Voice Cloning \cite{jemine2019master}. Then the participants crop a video file of length equal to the length of the audio fake. Finally, the audio fake is lip-synced onto the video clip using Wav2Lip. 
Participants used the above methodology and generated the Deepfakes highlighted in Table \ref{tab:table8}.

\begin{table}
\begin{center}
\caption{\label{tab:table8} (Output from pre-defined study) Based on human review, we find that $58.6\%$ of participants are able to create a Deepfake. Note that these numbers do not represent if the \df generated can be detected as a fake by the human examiners or software detector.}
\vspace{+1cm}
\begin{tabular}{||c c ||} 
 \hline
 Deepfake component created & Percentage \\ [0.5ex] 
 \hline\hline
 Deepfake audio and video & 58.6\% \\
 \hline
 Deepfake audio only & 31.0\% \\ 
 \hline 
 Neither audio nor video & 10.4\% \\
 \hline
\end{tabular}
\end{center}

\end{table}

\subsubsection{Deepfakes detection} \label{subsubsec:dfdetected2}
Since all the fakes were generated using the same tools, they have similar fingerprints introduced by the Deepfake generator. We ran the same detection techniques on all the fakes and the summary is as follows - 

\begin{enumerate}
  \item Detection algorithms provided by Deepware's API detected all videos as suspicious. Seferbekov algorithm reported a score of 95-99 (100 is fake), while Avatarify and Deepware algorithms reported score between 17-32. 
  \item {Human analyzers unanimously agreed that the videos with muted audio appeared realistic without any distortions
  }
  \item Human analyzers unanimously agreed that the fake audio without video sounded vaguely like Kevin 
  \item Human analyzers unanimously agreed that all videos were slightly out of sync in parts when examined very carefully.
  \vspace{-0.5cm}
\end{enumerate}

\section{Discussion and Conclusions} \label{sec:disussions}
 
In this paper, we conducted a systematic study on the accessibility of Deepfake creation technology. 

\textbf{{Deepfake awareness.}} Majority of our study participants (84.6\%) have heard and encountered Deepfakes on social media websites or video sharing platforms. This is a positive step as awareness is the first step in fighting the challenges posed by Deepfakes.

\textbf{{Deepfake generation is not difficult for a novice.}} A substantial number of novice users 58.6\% were able to generate Deepfakes successfully in a limited time frame, given appropriate tools. Even without guidance on which tools to be used, 23.1\% of the novice users were able to generate fakes. This ease in generation can be attributed to Deepfake generation code being open-sourced on Google Colaboratory and instructions being readily available.

\textbf{{Quality of Deepfakes created.}} We note that of the Deepfake videos created by participants successfully from the open-ended study, roughly $53.3\%$ were flagged as fake and another $26.7\%$ are flagged as suspicious by our chosen software detection tool. This implies that the Deepfake generated by novice users are not realistic enough to fool a software detector. We also use the input of a set of five human reviewers. Even in this case, we find that $100\%$ videos are flagged as fake based on a combined audio-visual inspection. However, from the pre-determined tool study, all the videos were detected as suspicious by the same detection tool. Even though $100\%$ of the videos were flagged as fake by human reviewers, none of them demonstrated any visual distortion in the video component unlike the outputs from the open ended study. Additionally, all audio fakes sounded vaguely like the target person, unlike 65.3\% which sounded like human but not vaguely like the target person in the open-ended study. Hence, we achieved better quality fakes in the pre-determined tool study, although we highlight that existing available deepfake creation systems still fall shot at fidelity.

\textbf{{Key challenges in generating fakes.}} Our study demonstrated that the key challenges are the identification of the right tool followed by debugging or using the tool. This can be demonstrated by the jump in the fakes created between the two studies. Additionally, there are many tutorials publicly available to help guide this journey.

We discuss limitations and future directions for our work in Appendix \ref{sec:appendix}.

\bibliographystyle{unsrt}
\bibliography{main.bib}

\appendix
\section{Appendix} \label{sec:appendix}

\subsection{Deep generative networks literature} \label{subsec:dfliterature}

The most popular generative model used in literature is Generative Adversarial Networks (GANs) \\ \cite{goodfellow2014generative}. The network consists of two parts - the first one being a Generator and the second one being a Discriminator. The task of the Generator is to take a low-dimensional noise vector and map it to a realistic looking ``unseen" image that matches the probability distribution of the training set. The task of the Discriminator is to successfully distinguish between ``fake" images generated by the Generator and real images from the training dataset. The task of the Generative model is to successfully fool the Discriminator into classifying its outputs as  ``real" images. In the process, the Generator \textit{learns} to produce more and more realistic-looking images. Furthermore, the fake images can be \textit{conditioned} to look like a reference input image 
, such as in Pix2Pix \cite{isola2017image}, which is a generative model that enables image to image transformation. 

Pix2Pix and CycleGAN \cite{zhu2017unpaired} are popularly used for \textit{style transfer} where the task is to transfer the style characteristics of the source image to the target image. 
Conditional GANs such as the Glow model \cite{kingma2018glow} allow a user to condition the output of the generator on user-defined parameters such as hair color, eye shape, and age among others. Glow also allows a user to morph seamlessly between two different face images. 

Note that deep generative networks have been used successfully for various applications such as producing realistic animation, image and video editing such as inpainting, denoising, 3-D model generation, lossless multi-media compression \cite{gui2021review} apart from Deepfake generation.

\subsection{Participant compensation} \label{subsec:partcomp}

Each candidate was compensated for their time and effort. For the open-ended Deepfake generation study, they were awarded 35 USD for their time and a 10 USD bonus if they are able to create a complete audio-video fake. For the pre-defined Deepfake generation study they were awarded 45 USD for their time and 10 USD bonus if they are able to complete audio-video fake. This compensation was structured to reward their time, increase motivation during the session and reduce dropouts. All the compensation was paid out in the form of gift cards from a leading e-commerce company. 

\subsection{Participant demographics} \label{subsec:partdemo}

\begin{figure}[!t]
\centering
\resizebox{\columnwidth}{!}{
\begin{tabular}{cc}
    \centering
    
\begin{tikzpicture}
\begin{axis}[
ybar,
x = 1.3cm,
enlargelimits=0.15,
symbolic x coords={Female,Male,Non-binary},
ylabel = Number of participants,
xlabel = (a) Participant gender,
xtick=data,
nodes near coords,
]
\addplot coordinates {(Female,21) (Male,16) (Non-binary,2)};
\addplot coordinates {(Female,16) (Male,11) (Non-binary,2)};
\legend{Open-ended, Pre-defined}
\end{axis}
\end{tikzpicture}

\hspace{-0.8cm}
& 

\begin{tikzpicture}
    \begin{axis}[
            ybar,
            x = 1.1cm,
            xlabel = (b) Participant age,
            symbolic x coords={18--20,21--23,24--26,27--29,30--39,40--49},
            xtick = data,
            nodes near coords,
        ]
        \addplot table[x=interval,y=study1]{\age};
        \addplot table[x=interval,y=study2]{\age};
        \legend{Open-ended, Pre-defined}
    \end{axis}
\end{tikzpicture} 

\end{tabular}  }

\resizebox{0.8\columnwidth}{!}{
\begin{tikzpicture}
\begin{axis}[
ybar,
x = 1.2cm,
ylabel = Number of participants,
xlabel = (c) Participant race,
symbolic x coords={White,Black,Indian,Chinese,Korean,Asian*,Hispanic},
xtick=data,
nodes near coords,
]
\addplot coordinates {(White,5) (Black,6) (Indian,11) (Chinese,9) (Korean,1) (Asian*,3) (Hispanic,4)};
\addplot coordinates {(White,4) (Black,3) (Indian,7) (Chinese,8) (Korean,1) (Asian*,3) (Hispanic,3)};
\legend{Open-ended, Pre-defined}
\end{axis}
\end{tikzpicture}}

    \caption{Demographic breakdown of participants, in terms of (a) gender, (b) age, (c) race (*Asians not counted under Indian, Chinese or Korean). We plot numbers corresponding to the first study (with open-ended objective, n=39 participants) in blue and those corresponding to the second study (pre-defined objective, n = 29 participants) in red. }
    \vspace{0.5cm}
    \label{fig:demo}
\end{figure}
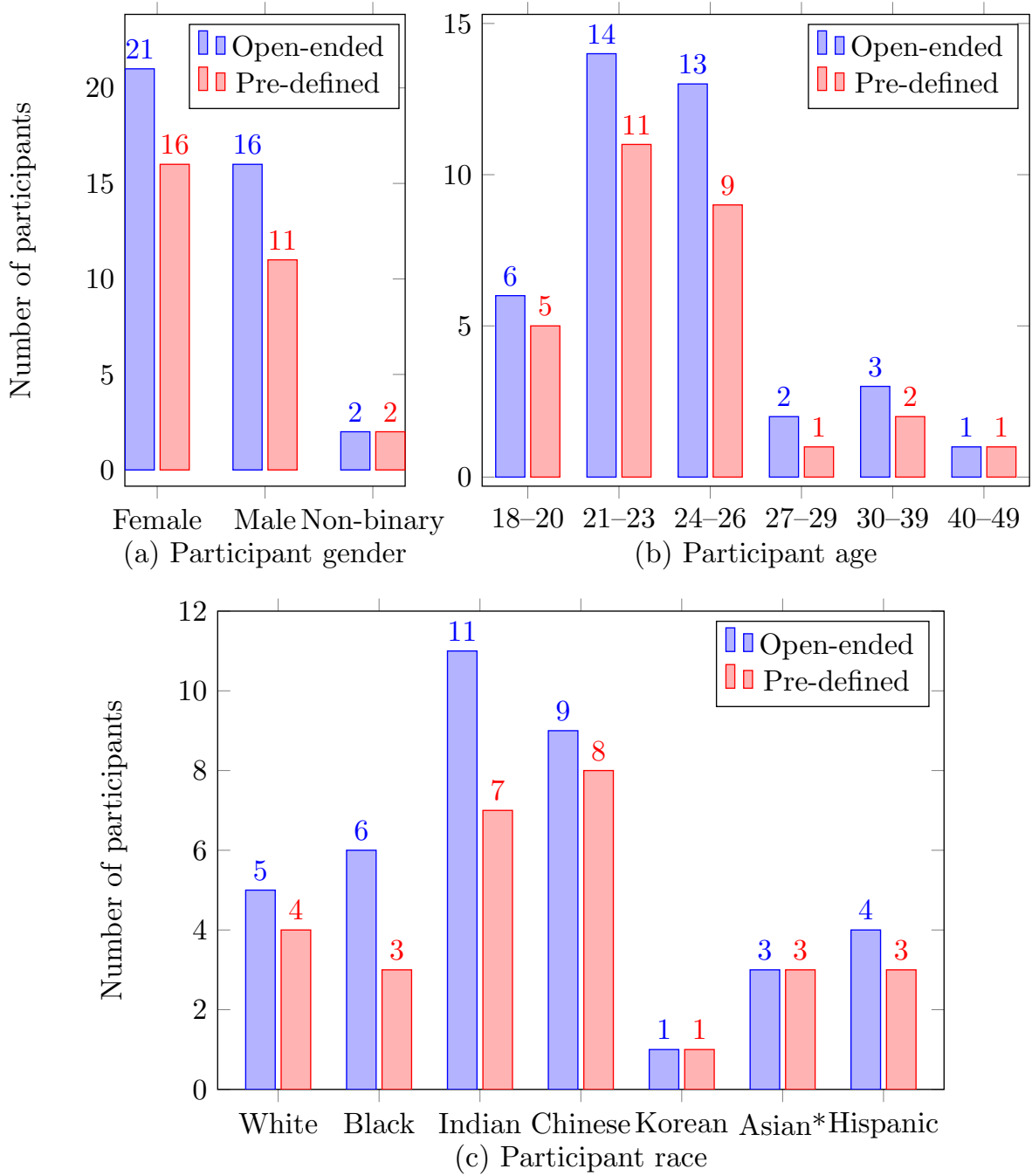

We chart out the demographics of the participants who actually participated in our study in terms of (a) gender (female, male, non-binary), (b) age range (18--20,21--23,24--26,27--29,30--39,40--49) and (c) race (White, Black or African American, Asian-Indian, Asian-Chinese, Asian-Korean, Asian-Other, and Hispanic). We plot numbers corresponding to the \textit{open-ended} study in blue and those corresponding to the \textit{pre-defined} objective in red and represent our data in Figure \ref{fig:demo}. We note that a total of $n=39$ participants were present for the first part of the study and $n=29$ participants were present for the second part. Breakdown of technical expertise for participants in both studies is also given in Table \ref{tab:table1} and \ref{tab:table7} respectively.

\pgfplotstableread[row sep=\\,col sep=&]{
    interval & study1 & study2 \\
    18--20     & 6  &  5 \\
    21--23     & 14 & 11  \\
    24--26    & 13 & 9  \\
    27--29   & 2 & 1   \\
    30--39   & 3  & 2  \\
    40--49  & 1  & 1 \\
    }\age
    
\subsection{Ethical consideration} \label{subsec:ethconsider}

This study was approved by the Institutional Review Board (IRB) of the first author's university. All participants were required to be over the age of 18 and provided  consent for the study. We provided necessary caution to all participants at the start and end of each of the studies that the objective of this research study was to understand the current state of Deepfakes from an academic perspective only. Participants were strongly suggested to use any knowledge gained about Deepfakes from this study judiciously. Participants were also directed to delete all the Deepfake-related files from their local computer or cloud after the session ended. 

\subsection{ Limitations} \label{subsec:limit}
A remote study helped us in understanding Deepfake generations while simultaneously ensuring health safety; however, such study design comes with several limitations. 
{Our participants sample represented young adults who are pursuing both undergraduate and graduate degrees}. We relied on participants self reported their coursework to recruit and assign groups. The media files provided of the target person had all the footage with the face looking straight into the camera. Such front-facing video is optimal for creating fakes and may not be always available when creating fakes in the real world. The participants were also limited by the capabilities of their Operating system, hardware and internet speed. Many Deepfake tools run only on Microsoft Windows, require GPU such as NVIDIA CUDA and require downloading files that could be as large as a few gigabytes. All these factors can limit the Deepfake generated by participants. 
The time-bound nature of the study is another factor that can inhibit the results. Many participants reported in the post survey that they would be able to create the fakes if more time was provided. Lastly, self-reporting surveys have biases such as social desirability biases.

\subsection{Future directions} \label{subsec:future}

Based on our study and conclusions, we propose some key steps to further the findings of our analysis. 

\textbf{Revising the parameters of our study:} For future, we propose to conduct this study with a larger sample set that is representative of a diverse population. One of the limitations of our study was also the time constraint under which participants had to create the Deepfake. Instead of two hours, one may increase the time frame of the study and infer the results. 

\textbf{Social impact and introducing restrictions:} Since we survey the impact of increased accessibility of Deepfake software, another aspect of the study may involve inspecting the implications of sharing Deepfakes online through social media websites. We can also investigate guidelines and restrictions on sharing such content online.

\textbf{Easier access:} The task of creating Deepfakes may be outsourced to external agencies or artists, instead of novice users. We can evaluate the quality of Deepfakes generated by experts under no time constraints and compare them against the data collected for the users in this study.

\section{Supplementary results} \label{sec:supplementary}

\begin{figure}[b]
\centering
\includegraphics[width=0.25\textwidth]{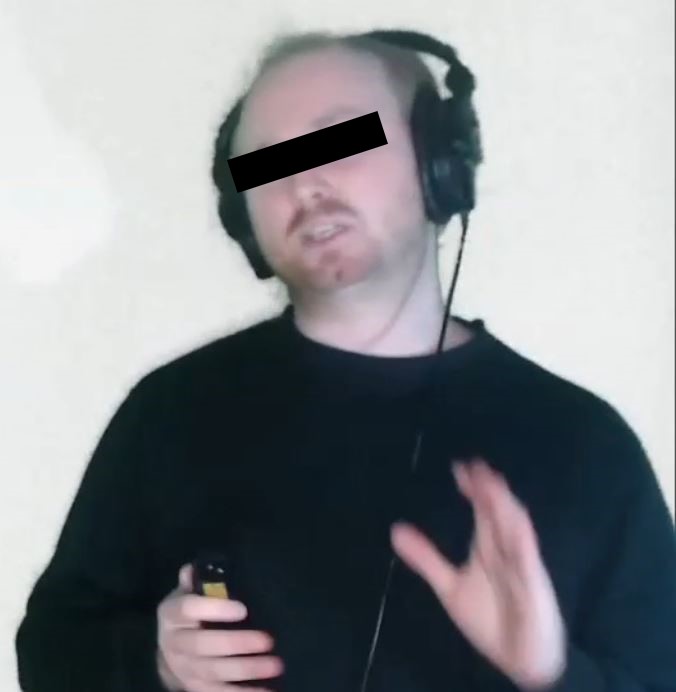}
\caption{A realistic Deepfake produced from study-1}\label{fig:fig5}
\end{figure}

\begin{figure}[b]
\centering
\includegraphics[width=0.25\textwidth]{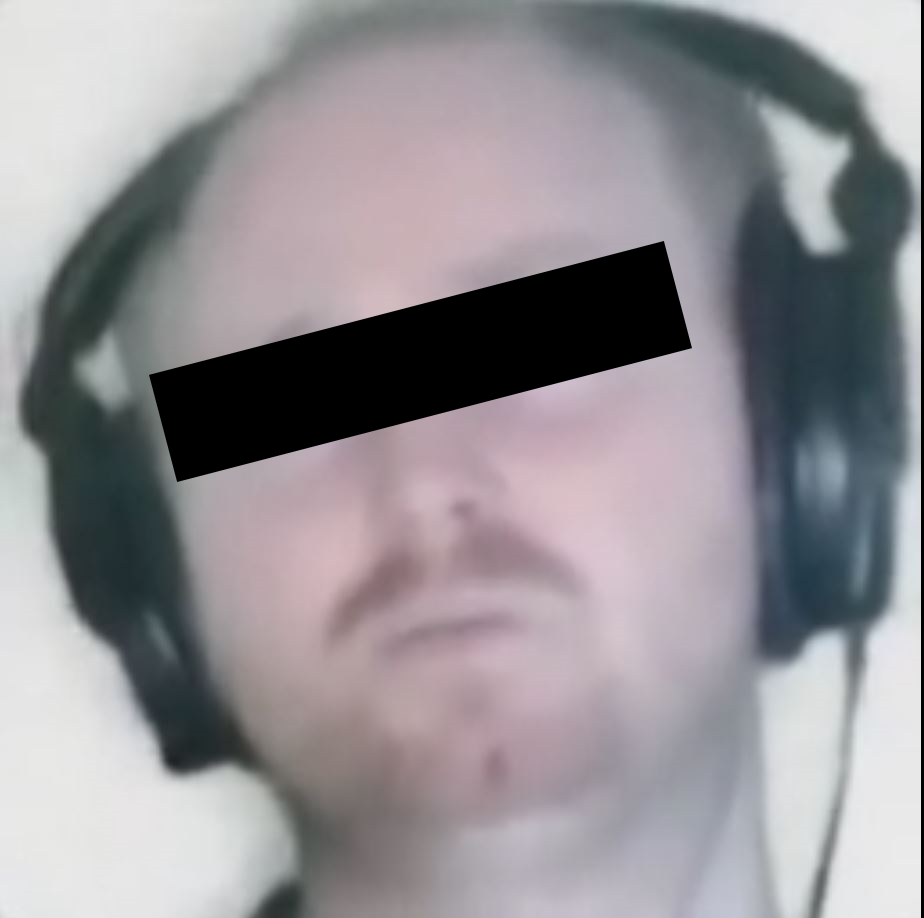}
\caption{A realistic Deepfake produced from study-2}\label{fig:fig6}
\end{figure}

\begin{figure}[b]
\centering
\includegraphics[width=0.25\textwidth]{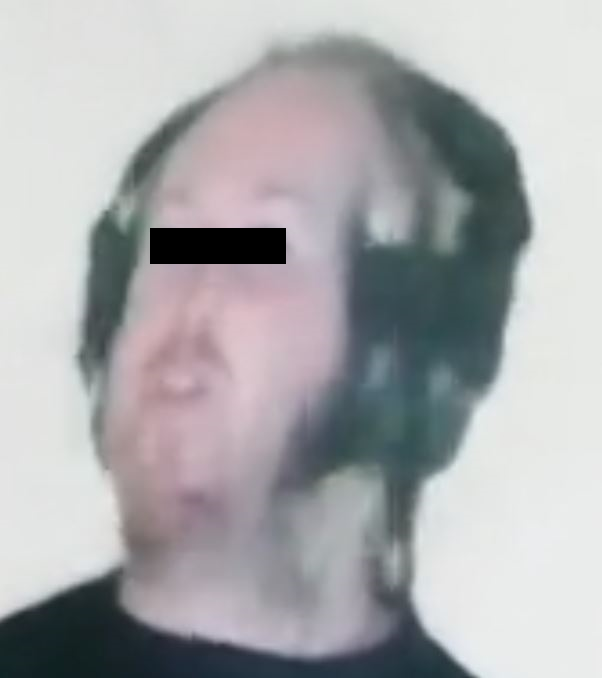}
\caption{A distorted Deepfake from study-1}\label{fig:fig7}
\end{figure}

\begin{figure}[b]
\centering
\includegraphics[width=0.25\textwidth]{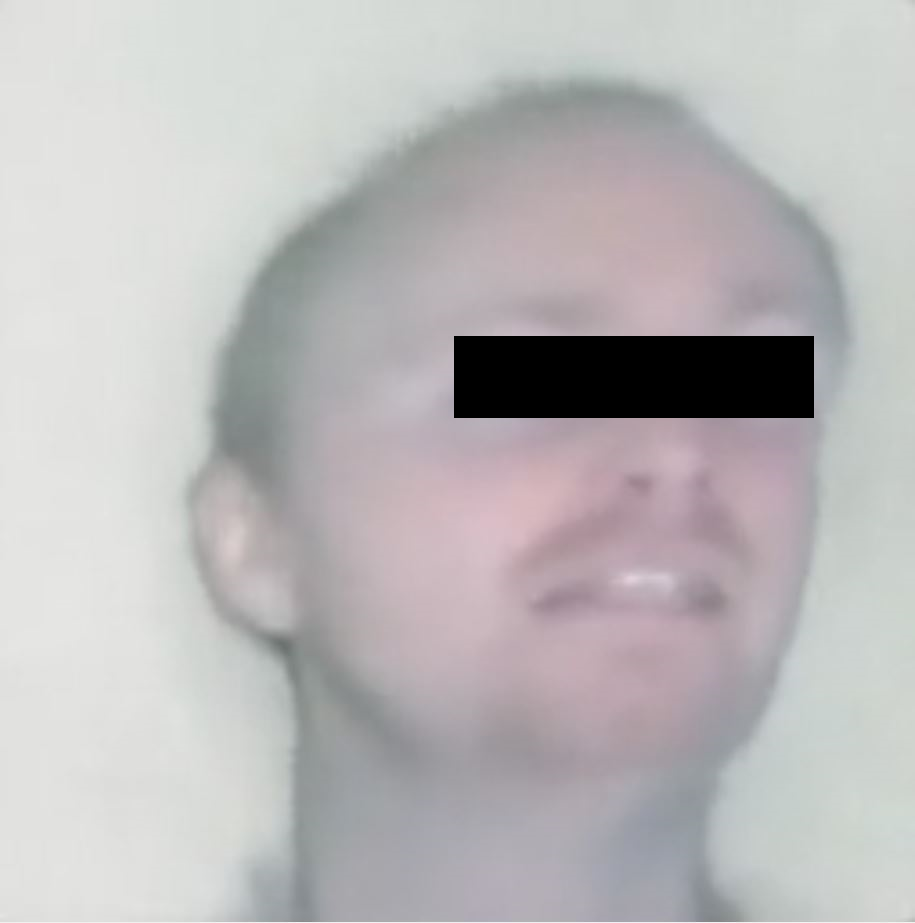}
\caption{A distorted Deepfake from study-2}\label{fig:fig8}
\end{figure}

The results obtained from both studies exhibited diverse variations. The output was contingent upon the techniques employed for image/video cropping, and, the Deepfake generation tool used. Thus, demonstrated significant disparities. The input media files had the target person facing the camera, which considerably facilitated the generation of Deepfakes. Participants adopted varied cropping strategies, with some confining themselves to the facial region as seen in fig. \ref{fig:fig5} and others encompassing the upper body as seen in fig \ref{fig:fig6}. Our anecdotal findings suggested that automated deepfake detectors exhibited greater proficiency in detecting videos containing only the face, rather than those encompassing the upper body. Moreover, we noted that the output generated from the first study displayed more distortions (see fig. \ref{fig:fig7}) in comparison to that of the second study (see fig. \ref{fig:fig8}). This discrepancy could be attributed to the restricted toolkit available in the latter study.

\subsection{Acknowledgments} \label{subsec:acknowledgement}

This work was supported in part by NSF 2016061.
\vspace{0.3cm}

\end{document}